\documentclass{emulateapj}

\begin{document}

\shortauthors{Luhman et al.}
\shorttitle{Brown Dwarf Disk}

\title{Discovery of a Planetary-Mass Brown Dwarf with a Circumstellar Disk}

\author{
K. L. Luhman\altaffilmark{1},
Luc\'\i a Adame\altaffilmark{2},
Paola D'Alessio\altaffilmark{3}, 
Nuria Calvet\altaffilmark{4}, 
Lee Hartmann\altaffilmark{4}, 
S. T. Megeath\altaffilmark{5}, 
\& G. G. Fazio\altaffilmark{5}}

\altaffiltext{1}{Department of Astronomy and Astrophysics,
The Pennsylvania State University, University Park, PA 16802;
kluhman@astro.psu.edu.}

\altaffiltext{2}{Instituto de Astronom\'\i a, UNAM, Apartado Postal 70-264, 
Ciudad Universitaria, 04510, M\'exico DF, M\'exico;
adamel@astroscu.unam.mx.}

\altaffiltext{3}{Centro de Radioastronom\'\i a y Astrof\'\i sica, UNAM,
Apartado Postal 72-3 (Xangari), 58089, Morelia, Michoac\'an, M\'exico;
p.dalessio@astrosmo.unam.mx.}

\altaffiltext{4}{Department of Astronomy, The University of Michigan, 
500 Church Street, 830 Dennison Building, Ann Arbor, Michigan 48109;
ncalvet, lhartm@umich.edu.}

\altaffiltext{5}{Harvard-Smithsonian Center for Astrophysics, 60 Garden St.,
Cambridge, MA 02138; tmegeath, gfazio@cfa.harvard.edu.}

\begin{abstract}

Using the {\it Hubble Space Telescope}, 
the 4~m Blanco telescope at the Cerro Tololo Inter-American Observatory,
and the {\it Spitzer Space Telescope},
we have performed deep imaging from 0.8 to 8~\micron\ of the southern
subcluster in the Chamaeleon~I star-forming region. In these data,
we have discovered an object, Cha~110913-773444, whose colors and magnitudes
are indicative of a very low-mass brown dwarf with a circumstellar disk.
In a near-infrared spectrum of this source obtained with the
Gemini Near-Infrared Spectrograph, the presence of strong steam absorption
confirms its late-type nature ($\gtrsim$M9.5) while 
the shapes of the $H$- and $K$-band continua and the strengths of the 
Na~I and K~I lines demonstrate that it is a young, pre-main-sequence object 
rather than a field dwarf.
A comparison of the bolometric luminosity of Cha~110913-773444 to 
the luminosities predicted by the evolutionary models of Chabrier \& Baraffe
and Burrows \& coworkers indicates a mass of $8^{+7}_{-3}$~$M_{\rm Jup}$,
placing it fully within the mass range observed for extrasolar planetary
companions ($M\lesssim15$~$M_{\rm Jup}$).
The spectral energy distribution of this object exhibits mid-infrared excess
emission at $\lambda>5$~\micron, which we have successfully modeled in terms
of an irradiated viscous accretion disk with 
$\dot{M}\lesssim10^{-12}$~$\rm M_{\odot}\,yr^{-1}$.
Cha~110913-773444 is now the least massive brown dwarf observed to have
a circumstellar disk, and indeed is one of the least massive 
free-floating objects found to date. These results 
demonstrate that the raw materials for planet formation exist around 
free-floating planetary-mass bodies.

\end{abstract}


\section{Introduction}
\label{sec:intro}

Planets are born from disks of gas and dust that surround newly formed stars. 
These disks exist around stars covering a wide range of masses and even have 
been found around objects at substellar masses \citep[e.g.,][]{com98}.
In fact, the least massive brown dwarf discovered 
with a disk \citep{luh05ots} 
has a mass near the upper limit of extrasolar planetary companions 
\citep[$M\sim15$~$M_{\rm Jup}$,][]{mar05}.
These measurements raise the question of whether even 
smaller brown dwarfs at planetary masses possess disks out of which planets 
might form. 

The Chamaeleon~I star-forming region is a prime location in which to 
search for planetary-mass brown dwarfs with disks. Because of its close 
proximity to the Sun \citep[$d=160$-170~pc,][]{whi97,wic98,ber99},
substellar members of this 
cluster have relatively bright apparent magnitudes. In addition, the cluster 
is compact enough that it can be mapped in a reasonable amount of observing 
time. Taking advantage of these attractive characteristics, we have 
obtained deep broad-band images of the southern subcluster in Cha~I from 0.8 to 
8~\micron\ with the {\it Hubble Space Telescope (HST)}, 
the 4~m Blanco telescope at the Cerro Tololo Inter-American Observatory, 
and the {\it Spitzer Space Telescope}.
In this letter, we describe these imaging observations, 
use the resulting photometry to identify a promising candidate brown dwarf 
with a disk, confirm it as a young brown dwarf with near-infrared (IR)
spectroscopy, estimate its extinction, luminosity, and mass, and 
compare its mid-IR excess emission to our model predictions for emission 
from a circumstellar disk.

\section{0.8-8~\micron\ Imaging of Chamaeleon I}
\label{sec:images}

We used the Advanced Camera for Surveys (ACS) aboard {\it HST} 
to image a $13\farcm3\times16\farcm7$ area centered at a 
$\alpha=11^{\rm h}07^{\rm m}45^{\rm s}$, $\delta=-77\arcdeg40\arcmin00\arcsec$
(J2000) during several dates in 2004 and 2005. 
ACS provided a plate scale of $0\farcs05$~pixel~$^{-1}$ and
a field of view of $3\farcm4\times3\farcm4$. 
The target area was imaged with a $4\times5$ map of 
contiguous fields of view.
At a given cell in the map, we obtained one 850~s exposure in the F775W 
filter (0.775~\micron) and one 350~s exposure in the F850LP filter 
(0.85~\micron) at each position in a 2-point dither pattern. 
The resulting images were processed and calibrated by automated software at 
the Space Telescope Science Institute. On the nights of 2004 April 28 and 29, 
we obtained images of a $20\arcmin\times20\arcmin$ area centered at 
$\alpha=11^{\rm h}07^{\rm m}45^{\rm s}$, $\delta=-77\arcdeg38\arcmin20\arcsec$
(J2000) with the Infrared Side Port Imager (ISPI) at the 4~m Blanco 
telescope at the Cerro Tololo Inter-American Observatory. ISPI provided 
a plate scale of $0\farcs3$~pixel~$^{-1}$ and a field of view of 
$10\farcm25\times10\farcm25$.
The target area was imaged with a $2\times2$ map of contiguous
fields of view through 
the $J$, $H$, and $K_s$ filters (1.25, 1.6, 2.2~\micron). 
At each cell in the map, short 
exposures were obtained at closely separated dither positions such that a 
total exposure time of 24~min was achieved in each filter. These data were 
processed using standard image reduction methods and were calibrated with 
photometry from the Two Micron All-Sky Survey. As a part of the Guaranteed 
Time Observations of the instrument team for the Infrared Array Camera 
\citep[IRAC;][]{faz04}, we obtained images of two overlapping 
$20\arcmin\times15\arcmin$ areas in Cha~I at 3.6, 4.5, 5.8, and 
8.0~\micron\ with IRAC aboard the {\it Spitzer Space Telescope} \citep{wer04}.
The details of these observations and the analysis of the resulting data are 
provided by \citet{luh05frac}.

\section{Identification of a Candidate Brown Dwarf with a Disk}

We measured photometry for all point sources appearing in the ACS, ISPI,
and IRAC images.
We then searched these data for objects exhibiting the colors and magnitudes 
expected of brown dwarfs with disks. This was done by first identifying 
objects that appeared to be cool and faint according to diagrams of 
$m_{775}$ versus $m_{775}-m_{850}$ \citep{luh05wfpc} 
and $m_{775}-K_s$ versus $J-H$ \citep{luh00}. For these candidate 
brown dwarfs, we checked the mid-IR colors $[3.6]-[4.5]$, $[4.5]-[5.8]$, 
and $[5.8]-[8.0]$ for values that were indicative of excess emission 
from disks \citep{luh05frac}. 
This process produced one promising candidate for a planetary-mass brown dwarf 
with a disk. The coordinates of the candidate are 
$\alpha=11^{\rm h}09^{\rm m}13.63^{\rm s}$, 
$\delta=-77\arcdeg34\arcmin44\farcs6$
(J2000) and we assign to it the name Cha~110913-773444 (hereafter 
Cha~1109-7734). For this object, we measured photometry of $m_{775}=23.19$, 
$m_{850}=21.59$, $J=17.45$, $H=16.34$, $K_s=15.61$, 
$[3.6]=14.70$, $[4.5]=14.38$, $[5.8]=14.11$, and $[8.0]=13.49$.  
The errors are 0.02 and 0.04 for the former seven and the latter two 
measurements, respectively. 

\section{Spectral Classification}

\begin{figure}
\epsscale{0.5}
\plotone{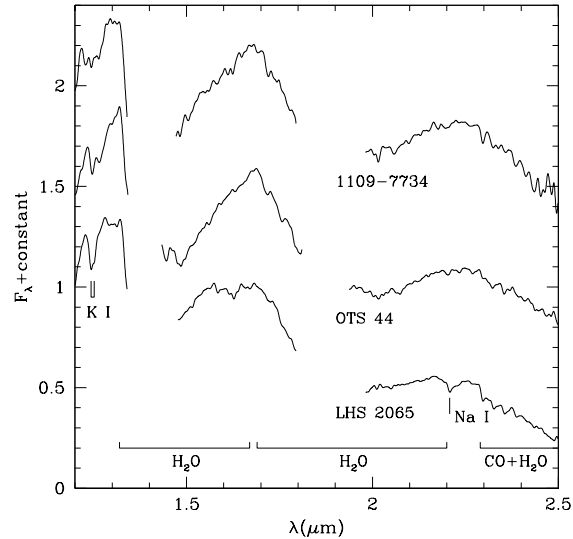}
\caption{ 
GNIRS spectrum of the candidate young brown dwarf Cha~1109-7734 compared 
to spectra of the known young brown dwarf OTS~44 and the old cool dwarf 
LHS~2065 (M9V). Like the latter two objects, Cha~1109-7734 
exhibits broad, deep absorption in H$_2$O, demonstrating that it has a late
spectral type. The weak K~I and Na~I absorption lines and the triangular 
shape of the continuum between 1.5 and 1.8~\micron\ in the spectrum of 
Cha~1109-7734 indicate a low surface gravity, and hence young age, like that 
of OTS~44. The spectra are displayed at a resolution of $R=200$ and are
normalized at 1.68~\micron.}
\label{fig:spec}
\end{figure}

To determine if Cha~1109-7734 is a young brown dwarf, we observed it 
spectroscopically with the Gemini Near-Infrared Spectrograph (GNIRS) at Gemini 
South Observatory during the nights of 2005 March 23 and 25. The observing and 
analysis procedures were the same as those used for the Chamaeleon brown dwarf 
OTS~44 \citep{luh04ots}. The resulting spectrum of Cha~1109-7734 is shown in 
Figure~\ref{fig:spec}. 
The spectrum exhibits strong H$_2$O absorption bands, demonstrating 
that Cha~1109-7734 is a cool object rather than an early-type field star 
or an extragalactic source. Both young brown dwarfs and low-mass field stars 
have cool atmospheres, but their surface gravities differ significantly. 
Therefore, we can distinguish between these two possibilities for 
Cha~1109-7734 by examining spectral features that are sensitive to gravity, 
such as the shapes of the $H$- and $K$-band continua that are induced by the 
H$_2$O absorption bands and the strengths of Na~I and K~I absorption
lines \citep{luh98,luc01,gor03,mc04}.
To do this, we compare in Figure~\ref{fig:spec} 
the spectrum of Cha~1109-7734 to GNIRS 
data for the young brown dwarf OTS~44 and the cool field dwarf LHS~2065. 
Cha~1109-7734 has a sharply peaked, triangular continuum and weak Na~I 
and K~I lines like OTS~44 rather than the broad plateau and strong absorption 
lines that characterize LHS~2065. Based on this comparison, we conclude that 
Cha~1109-7734 is a young member of Cha~I rather than a field dwarf 
in the foreground or the background of the cluster. Further evidence of the 
youth and membership of Cha~1109-7734 will be provided later in this work 
when we demonstrate that it has a disk. In addition to surface gravity, the 
spectral type of Cha~1109-7734 is constrained by the H$_2$O absorption bands. 
The strengths of these bands for Cha~1109-7734 are similar to those of 
OTS~44, which was classified as $\gtrsim$M9.5 through a comparison to an 
optically-classified young brown dwarf \citep{luh04ots}. 
We therefore apply this spectral 
type to Cha~1109-7734 as well. Because the variation of H$_2$O absorption with 
optical spectral type is unknown for young objects later than M9, we can place 
only a limit on the spectral type.

\section{Extinction, Luminosity, and Mass}

Now that Cha~1109-7734 has been spectroscopically confirmed as a young 
brown dwarf, we estimate its extinction, luminosity, and mass. By comparing 
the near-IR spectrum and colors of Cha~1109-7734 to those of the 
known brown dwarfs OTS~44 \citep[$A_J=0.3$,][]{luh04ots}
and KPNO~4 \citep[$A_J=0$,][]{bri02}, we
derive an extinction of $A_J=0.3\pm0.3$. We estimate a bolometric luminosity of 
log~$L_{\rm bol}=-3.22$ for Cha~1109-7734 by combining our measurement of $H$, 
a distance modulus of 6.05, an $H$-band bolometric correction of BC$_H=2.7$
\citep{leg01,rei01,dah02}, and an absolute bolometric magnitude for the 
Sun of $M_{\rm bol \odot}=4.75$.
The combined uncertainties in $A_H$, $H$, BC$_H$, and the distance 
modulus ($\sigma\sim0.2$, 0.02, 0.2, 0.13)
correspond to an uncertainty of $\pm0.12$ in log~$L_{\rm bol}$. 

\begin{figure}
\epsscale{0.5}
\plotone{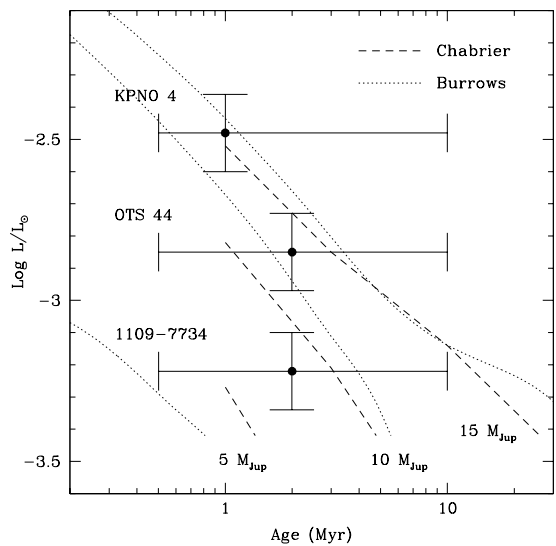}
\caption{ 
The luminosities of the young brown dwarfs KPNO~4, OTS~44, and
Cha~1109-7734 ({\it points; top to bottom}) are compared to the luminosities 
as a function of age predicted by the theoretical evolutionary models of 
\citet{cha00} ({\it dashed line}) and \citet{bur97} ({\it dotted line})
for masses of 5, 10, and 15~$M_{\rm Jup}$.
Cha~1109-7734 has a mass of $8^{+7}_{-3}$~$M_{\rm Jup}$ according to these
models.
}
\label{fig:lbol}
\end{figure}

We convert the luminosity estimate to a mass using theoretical relationships 
between luminosity, mass, and age. For the age of Cha~1109-7734, we adopt 
the median value of 2~Myr exhibited by the known members of Cha~I 
\citep{luh04cha}.  These members are spread across a range of luminosities 
at a given temperature, which could reflect either a range of ages 
($\tau\sim0.5$-10~Myr) or other 
phenomena, such as extinction uncertainties, unresolved binaries, and 
variability. Regardless of the source of this luminosity spread, we can 
account for it in our mass estimate by adopting lower and upper limits of 0.5 
to 10~Myr for the age of Cha~1109-7734. In diagrams of luminosity versus 
age in Figure~\ref{fig:lbol},
we plot Cha~1109-7734 with the luminosities predicted 
as a function of age for masses of 5, 10 and 15~$M_{\rm Jup}$ 
by \citet{bur97} and \citet{cha00}.  
Cha~1109-7734 has a mass of $8^{+7}_{-3}$~$M_{\rm Jup}$ according to both 
sets of evolutionary models. Mass estimates of this kind are prone to
systematic errors within the models \citep{bar02}, 
but these errors do not appear to be large if the models are used properly
\citep{luh05abdor1,luh05abdor2}. 
Thus, considering that the observed upper limit of extrasolar 
planetary companions is near 15~$M_{\rm Jup}$ \citep{mar05},
the mass of Cha~1109-7734 is very likely in the planetary regime. 
For comparison, we also include in Figure~\ref{fig:lbol} 
the young brown dwarfs KPNO~4 and OTS~44. We have revised the 
previous luminosities of these objects \citep{bri02,luh04ots}
using the values of BC$_H$ and $M_{\rm bol \odot}$ adopted in this work 
for Cha~1109-7734, arriving at log~$L_{\rm bol}=-2.48$ and -2.85 
for KPNO~4 and OTS~44, respectively. Cha~1109-7734 is more than twice as faint
as OTS~44, which is the least massive brown dwarf known to have a disk prior 
to this study \citep{luh05ots}.

\section{Infrared Excess Emission}

\begin{figure}
\epsscale{0.5}
\plotone{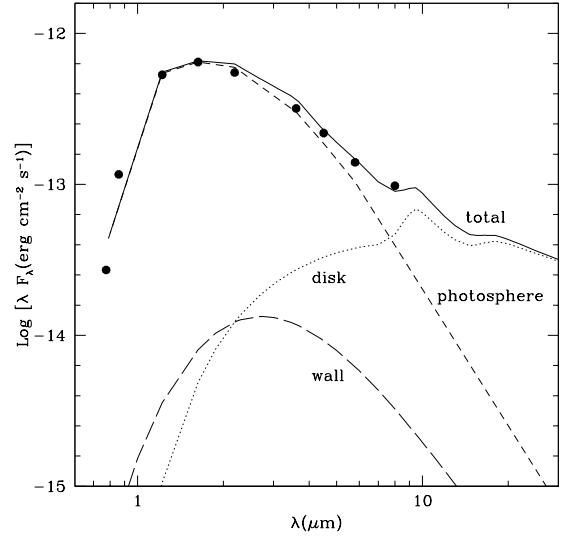}
\caption{ 
Spectral energy distribution of the young brown dwarf Cha~1109-7734 
({\it points}). Relative to the distribution 
expected for its photosphere ({\it short dashed line}), this brown dwarf 
exhibits significant excess emission at wavelengths longer than 
5~\micron. The excess flux is modeled in terms of emission from a circumstellar
accretion disk ({\it dotted line}) and a
vertical wall at the inner disk edge ($R_{\rm wall}=2.1$~$R_*$, 
$H_{\rm wall}=0.17$~$R_*$, {\it long dashed line}).
The sum of this disk model and the 
photosphere ({\it solid line}) is a reasonable match to the data for 
Cha~1109-7734.
}
\label{fig:sed}
\end{figure}

Cha~1109-7734 was originally identified as an object that might have a 
disk on the basis of mid-IR colors that were suggestive of excess 
emission from cool dust. We now examine this evidence of a disk in more detail. 
As done in our previous analysis of the disk-bearing brown dwarf OTS~44 
\citep{luh05ots},
we use the young brown dwarf KPNO~4 to represent the stellar photosphere 
of Cha~1109-7734. In Figure~\ref{fig:sed}, 
we plot the spectral energy distribution (SED), $\lambda\times F_{\lambda}$, 
for Cha~1109-7734 after an extinction correction of $A_J=0.3$. 
The SED of KPNO~4 is scaled to match the $H$-band flux of Cha~1109-7734 
and is included in Figure~\ref{fig:sed}. 
As in the case of OTS~44 \citep{luh05ots}, the fluxes 
of Cha~1109-7734 are photospheric in the optical and near-IR bands, while 
significant excess emission is present at wavelengths longer than 5~\micron.  
We compared this excess emission from Cha~1109-7734
with the predictions of a model of an 
irradiated accretion disk. For these calculations, we adopted a stellar mass 
of $M_*=8$~$M_{\rm Jup}$, a stellar radius of $R_*=1.8$~$R_{\rm Jup}$, 
a uniform grain size distribution characterized by the standard power 
law $n(a) \sim a^{-3.5}$ with minimum and maximum grain sizes 
$a_{min}=0.005$~\micron\ and $a_{max}=0.25$~\micron\ \citep{mat77},
a uniform mass accretion rate, and a disk inclination of $i=15\arcdeg$. 
Standard disk models include a vertical 
wall at the dust destruction temperature of 1400~K. However, because of the 
low luminosity of Cha~1109-7734, the dust destruction radius is 
$\sim1.8$~$R_*$.  
Under the assumption that magnetospheric accretion is taking place in this 
object, we have truncated the disk and placed the wall at a typical 
magnetospheric radius of $\sim2.1$~$R_*$ \citep{muz05}, 
which corresponds to a temperature of 1300~K. For the height of 
the disk wall we have used $H_{\rm wall}=0.17$~$R_*$, which was derived from
equations in \citet{muz04}. Descriptions of the method used 
to calculate the disk structure and emergent intensity are provided 
in our previous work \citep{dal98,dal99,dal01,dal05,muz03}.
We find that models with accretion rates of 
$\dot{M}>10^{-12}$~$\rm M_{\odot}\,yr^{-1}$
produce too much mid-IR excess emission. Meanwhile, we can place only 
an upper limit of $\dot{M}\lesssim10^{-12}$~$\rm M_{\odot}\,yr^{-1}$
on the accretion rate because irradiation rather than accretion dominates the 
heating of the inner disk for $\dot{M}<10^{-12}$~$\rm M_{\odot}\,yr^{-1}$.
In Figure~\ref{fig:sed}, we show the disk SED produced by the model with 
$\dot{M}=10^{-12}$~$\rm M_{\odot}\,yr^{-1}$, which fits the observed
excess emission reasonably well. 


We have presented the discovery of the faintest brown dwarf known to harbor 
a circumstellar disk. Our best estimate of the mass of this object 
($M=8^{+7}_{-3}$~$M_{\rm Jup}$) places it within the mass range of 
extrasolar planetary companions ($M<15$~$M_{\rm Jup}$). 
Thus, the basic ingredients for making planets are 
present around free-floating planetary-mass bodies. Additional observational 
and theoretical work is needed to determine if and how planets form in these 
disks.

\acknowledgements

We acknowledge support from grant NAG5-11627 from the NASA Long-Term Space
Astrophysics program and grant GO-10138 from the Space Telescope Science 
Institute (K. L.), grant 172854 from CONACyT (L. A.), 
grants from CONACyT and PAPIIT/DGAPA, M\'exico (P. D.), and NASA
grants NAG5-9670 and NAG5-13210 (N. C., L. H.). 
This work is based on observations made 
with the {\it Spitzer Space Telescope}, the {\it HST}, Cerro 
Tololo Inter-American Observatory, and Gemini Observatory. {\it Spitzer} is 
operated by the Jet Propulsion Laboratory at the California Institute of 
Technology under NASA contract 1407.
Support for the IRAC instrument was provided by NASA through contract 
960541 issued by JPL.
The {\it HST} observations were obtained at the Space Telescope 
Science Institute, which is operated by the Association of Universities for 
Research in Astronomy, Inc.\ (AURA), under NASA contract NAS 5-26555. 
Cerro Tololo Inter-American Observatory is operated by AURA under a 
contract with the NSF. Gemini Observatory is operated by AURA under a
cooperative agreement with the NSF on behalf of the Gemini partnership: the 
National Science Foundation (United States), the Particle Physics and Astronomy 
Research Council (United Kingdom), the National Research Council (Canada),
CONICYT (Chile), the Australian Research Council (Australia), CNPq (Brazil) 
and CONICET (Argentina).

\end{document}